\documentclass[sigconf]{acmart}

\usepackage{todonotes}
\usepackage{pgfplots}
\usepackage{pgfplotstable}
\pgfplotsset{compat=1.7}

\newcommand{\rqa}{(RQ1) On what basis does the development team prioritize which groups of crash reports to investigate first?}
\newcommand{\rqb}{(RQ2) How effective is the information from the grouping and ranking approaches in assisting the development team in fixing bugs?}
\newcommand{\rqc}{(RQ3) What additional types of information are useful to developers in the bug-fixing process?}

\AtBeginDocument{%
  \providecommand\BibTeX{{%
    \normalfont B\kern-0.5em{\scshape i\kern-0.25em b}\kern-0.8em\TeX}}}

\setcopyright{acmcopyright}
\copyrightyear{2024}
\acmYear{2024}
\setcopyright{acmlicensed}\acmConference[ICSE-SEIP '24]{46th International Conference on Software Engineering: Software Engineering in Practice}{April 14--20, 2024}{Lisbon, Portugal}
\acmBooktitle{46th International Conference on Software Engineering: Software Engineering in Practice (ICSE-SEIP '24), April 14--20, 2024, Lisbon, Portugal}
\acmDOI{10.1145/3639477.3639730}
\acmISBN{979-8-4007-0501-4/24/04}

\begin{document}

\title{The Impact Of Bug Localization Based on Crash Report Mining: A Developers' Perspective}


\author{Marcos Medeiros}
\email{marcosamm@gmail.com}
\affiliation{%
  \institution{Federal University of Rio Grande do Norte}
  \city{Natal}
  \country{Brazil}
}

\author{Uirá Kulesza}
\email{uira@dimap.ufrn.br}
\affiliation{%
  \institution{Federal University of Rio Grande do Norte}
  \city{Natal}
  \country{Brazil}
}

\author{Roberta Coelho}
\email{roberta@dimap.ufrn.br}
\affiliation{%
  \institution{Federal University of Rio Grande do Norte}
  \city{Natal}
  \country{Brazil}
}

\author{Rodrigo Bonif\'{a}cio}
\email{rbonifacio@unb.br}
\affiliation{%
  \institution{University of Brasília}
  \city{Brasília}
  \country{Brazil}
}

\author{Christoph Treude}
\email{christoph.treude@unimelb.edu.au}
\affiliation{%
  \institution{University of Melbourne}
  \city{Melbourne}
  \country{Australia}
}

\author{Eiji Adachi}
\email{eijiadachi@imd.ufrn.br}
\affiliation{%
  \institution{Federal University of Rio Grande do Norte}
  \city{Natal}
  \country{Brazil}
}


\renewcommand{\shortauthors}{Medeiros and Kulesza, et al.}

\begin{abstract}
Developers often use crash reports to understand the root cause of bugs. However, locating the buggy source code snippet from such information is a challenging task, mainly when the log database contains many crash reports. To mitigate this issue,  recent research has proposed and evaluated approaches for grouping crash report data and using stack trace information to locate bugs. The effectiveness of such approaches has been evaluated by mainly comparing the candidate buggy code snippets with the actual changed code in bug-fix commits---which happens in the context of retrospective repository mining studies. Therefore, the existing literature still lacks discussing the use of such approaches in the daily life of a software company, which could explain the developers’ perceptions on the use of these approaches. In this paper, we report our experience of using an approach for grouping crash reports and finding buggy code on a weekly basis for 18 months, within three development teams in a software company. We grouped over 750,000 crash reports, opened over 130 issues, and collected feedback from 18 developers and team leaders. Among other results, we observe that the amount of system logs related to a crash report group is not the only criteria developers use to choose a candidate bug to be analyzed. Instead, other factors were considered, such as the need to deliver customer-prioritized features and the difficulty of solving complex crash reports (e.g., architectural debts), to cite some. The approach investigated in this study correctly suggested the buggy file most of the time---the approach's precision was around 80\%. In this study, the developers also shared their perspectives on the usefulness of the suspicious files and methods extracted from crash reports to fix related bugs.
\end{abstract}

\ccsdesc[500]{Software and its engineering~Maintaining software}

\keywords{Software crash, Bug correlation, Bug localization, Crash reports, Stack traces, Bug prioritization}


\received{20 February 2007}
\received[revised]{12 March 2009}
\received[accepted]{5 June 2009}

\maketitle

\section{Introduction}

Software developers often use crash reports to have an understanding of the system behavior in execution at the crash time, to identify and correct existing bugs, and to improve software quality \cite{An2015, ken, Laura, bettenburg2008makes, schroter2010stack}. Despite their usefulness, handling a substantial volume of crash reports can be challenging \cite{An2015, kinshumann2011debugging}. To address this issue, developers can group crash reports based on their similarity, for instance, which might reduce the efforts needed to identify and fix bugs. Indeed, several studies have explored stack trace aggregation for crash reports ~\cite{podgurski2003automated, khomh2011entropy, kim2011crashes, dang2012rebucket, Wang2013, Wang2016}, as well as its utility in bug localization and fixing~\cite{ball2003symptom, jones2002visualization, jones2005empirical, nessa2008software, schroter2010stack, wong2014boosting, gu2019does, Wu, wu2018changelocator}.

Although the techniques have evolved, most of the existing literature on bug localization based on crash report mining relies on empirical retrospective studies~\cite{khomh2011entropy, kim2011crashes, nessa2008software, Wang2013, Wang2016, schroter2010stack, wong2014boosting, Wu, wu2018changelocator} that analyze past system data to understand the performance of the designed techniques using open-source software data sets. Although it can be an adequate approach to analyze and compare the performance of existing techniques for grouping stack traces, it still does not explore the effective use of these approaches in an industrial context. Only a few studies report challenges and results of the practical application of such techniques in day-to-day activities in software development companies. 
Jarman et al.~\cite{jarman2021legion} report their experience applying a bug localization solution (Legion, a BugLocator extension) on Adobe Analytics repositories. Li and colleagues~\cite{li2022empirical} show the implications and challenges of using Information Retrieval-based Bug Localization (IRBL) techniques on 10 Huawei projects. Both papers mention the lack of studies applied in industry. This issue might be due to the difficulty in adapting existing techniques to the particular context of a company or due to the lack of configurable tools that practitioners could easily use on their projects.

In previous work \cite{medeiros2020}, we reported our experience applying a bug localization approach based on crash report mining in a retrospective dataset (without the involvement of the developer teams). We compared the changed files associated with a closed issue with the suggested buggy files returned by the approach. In this paper, we extend our previous work and present the results of using our approach with the development teams of a software company in the context of large-scale web-based systems implemented using Java Enterprise technologies. We conducted an industrial study to answer three research questions: 

\begin{itemize}
\item \emph{\rqa}
\item \emph{\rqb}
\item \emph{\rqc}
\end{itemize}

The remainder of this paper is organized as follows. Section \ref{sec:bug_location_approach} presents the bug localization approach. Section \ref{sec:empirical_study} characterizes the target systems and development teams and describes our study procedures. Section \ref{sec:results_and_discussion} provides a discussion of our results. Section \ref{sec:lessons_learned} shares learned lessons and implications for practitioners and researchers. Section \ref{sec:threats_to_validity} comments on threats to validity associated with this study. Section \ref{sec:related_work} discusses related work. Finally, Section \ref{sec:conclusion} concludes the paper and outlines directions for future work.

\section{Bug Localization Approach} \label{sec:bug_location_approach}

In this paper, we report our experience on using an approach that we customized and implemented based on previous work~\cite{Wu,Wang2016, medeiros2020}. The approach aims: (i) to group crash reports that generate similar stack traces; and (ii) to rank files suspected of causing crashes. The main goal is to facilitate the resolution of bugs associated with crash reports. The crash reports are clustered by stack traces according to specific rules (see details in Section \ref{sec:crash_report_grouping}). Also, a list of suspicious buggy files is generated based on the stack traces of the grouped crash reports (see details in Section \ref{sec:ranking_suspicious_files}).

\begin{figure}[htbp]
\centerline{
    \includegraphics[width=\columnwidth]{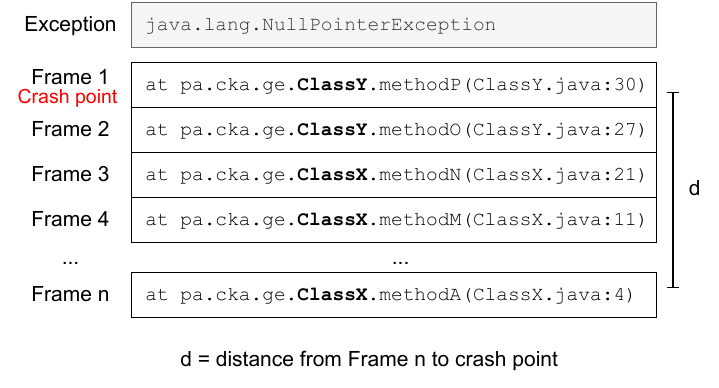}
}
\caption{Example of a Stack Trace}
\label{fig:crash_stack}
\end{figure}

A stack trace is an ordered set of frames $<F_1, F_2, ... F_n>$. Each frame $F_i$ consists of the qualified name of a method (a 3-tuple comprising the package, class, and method name), a filename (fileName), and a line number (line). $F_i = qMethodName|(fileName|:|line)$, where $i \in {1..n}$ is the frame position $F_i$ in the stack trace. The last executed frame (the most recent) is at the top, and the first (oldest) is at the bottom. The \emph{crash point} corresponds to the frame at the top of the stack, and the name of the file on that same line is called Top Frame File. \emph{Crash point} is also known as \emph{signaler} due to being where the exception is thrown. Fig.~\ref{fig:crash_stack} illustrates an example of a stack trace.

\subsection{Crash Report Grouping}\label{sec:crash_report_grouping}

We group crash reports using information available in the stack traces. The primary purpose is to aggregate strongly correlated crash reports, thus reducing software developers' cognitive effort in finding the most likely causes of a bug. 
In this work, we use the first four clustering levels from previous work~\cite{medeiros2020}.  
Next, we discuss the different grouping levels. We use the acronyms STA and STB to represent two stack traces in the crash report database.

{\bf (Level 1) Identical Stack Trace}. Initially, we group crash reports whenever two stack traces (STA and STB) are identical (STA = STB). The signature that represents each group is the stack trace itself. 

{\bf (Level 2) Equivalent Signature}. Even after grouping identical stack traces, it is possible to identify distinct groups whose signatures are almost identical. In these cases, the groups differ only in fine-grained attributes, such as the automatically generated accessor methods, among others. Consider two stack traces STA and STB, where 

\begin{itemize}
    \item \textit{at ...GeneratedMethodAccessor10184.invoke()} $\in$ STA 
    \item \textit{at ...GeneratedMethodAccessor10272.invoke()} $\in$ STB
\end{itemize}

\noindent That is, these lines of the stack trace only differ concerning the numbers \emph{10184} and \emph{10272}. For these small differences, we also consider two stack traces (e.g., STA and STB) equivalent and group their crash reports. 

Following a cumulative strategy, that is, joining previously created groups, the next levels are based on the rules proposed by Wang et al.~\cite{Wang2016}.

{\bf (Level 3) Crash Type Signature}. The rule here identifies similarities among correlated Level2 groups when comparing their crash type signatures (equivalent stack traces), and groups two stack traces STA and STB when one contains the other (STA $\subseteq$ STB or STB $\subseteq$ STA). To apply this rule, we only consider the rows of a stack trace with packages, classes, and methods (e.g., at java.lang.reflect.Method.invoke (Method.java:606)), since the rest might differ because of the idiom, for example. We also ignore the line number present in the stack trace because simple file formatting, blank line insertion, or comment inclusion can change the line number without modifying any code statement. 

{\bf (Level 4) Top Frame File}. This rule correlates Level3 groups when they have the same qualified file name in the crash point (see \textit{Frame 1} on Fig. \ref{fig:crash_stack}). Consider two stack traces STA and STB, where:
\begin{itemize}
    \item \textit{at s.p.ClassMBean.methodA(ClassMBean.java:280)} is the signaler of STA 
    \item \textit{at s.p.ClassMBean.methodB(ClassMBean.java:251)} is the signaler of STB
\end{itemize}
In both cases, the qualified file name is the same (\textit{s.p.ClassMBean}), so the rule groups this kind of similar stack traces.

\subsection{Suspicious file and method ranking}\label{sec:ranking_suspicious_files}

After grouping the crash reports using the 
procedures detailed in Section~\ref{sec:crash_report_grouping}, we mine information of the files present in the stack traces of the crash report groups. We use this information to rank the files that are more likely to cause the crashes in a given crash report group. To this end, our approach leverages the approach proposed by Wu et al. \cite{Wu}, ranking files instead of methods. This approach uses three criteria to perform the ranking: (i) Inverse Average Distance to Crash Point (IAD)---if a file appears closer to the crash point, it is more likely to cause the crash; (ii) Inverse Bucket Frequency (IBF)---if a file appears in stack traces caused by many different faults, it is less likely to be the cause of a specific fault; and (iii) File Frequency (FF)---if a file often appears in stack traces caused by a particular fault, it is likely to be the cause of this fault. 










The ranking of the suspicious files is accomplished using a combination of the three factors, generating a score for each file $f$ present in the stack traces of group $B$:
\begin{equation}
Score(f) = IAD(f,B)*IBF(f)*FF(f,B)\label{Score}
\end{equation}

This approach assigns higher scores to files that appear more often in the stack traces of a group, less frequently in stack traces from other groups, and that are closer to the crash point. The rank of suspected files containing crashes that triggered the crash report group is generated by calculating the score of each file in the stack traces and sorting them in descending order. 

{\bf \textit{Suggesting suspicious methods}}. For ranking methods, we use a different strategy. After ranking the suspicious files as presented above, we identify the methods (defined in such files) that appear in the stack traces of the respective group of crash reports and suggest them to the developers, showing the most frequent ones first.

Refer to our previous work~\cite{medeiros2020} for more details on clustering and ranking approaches.


\section{Empirical Study} \label{sec:empirical_study}

This study explores the benefits and challenges of applying crash report clustering and suspicious file/methods ranking approaches in a practical industrial scenario, considering the developers' perspective. We also analyzed how they use the information these approaches provide for classifying, identifying, and fixing bugs. To this end, we addressed the following research questions:

\emph{\rqa} --- This research question investigates what factors the development team considers when prioritizing the groups of crash reports they are interested in analyzing and solving the associated bugs.

\emph{\rqb} --- This research question investigates whether the bug is fixed in the classes and methods the approach suggests. We also examined whether the classes and methods ranking helped the development team to fix the bugs.

\emph{\rqc} --- This research question investigates whether additional information, such as affected users, affected URIs, stack trace samples, and crash report ID samples, among others, contributes to fixing bugs.

\subsection{Target Systems and Development Team}

We conducted our study in cooperation with a software development department of a public institution responsible for developing, maintaining, and evolving several large-scale Web-based systems. We focused on three specific software projects that maintain web systems that, together, receive more than one million daily requests. The systems contain approximately 3.5 million lines of code and twenty thousand classes. They are implemented using several Java Enterprise technologies, such as Spring, Java Server Faces, and Hibernate. Overall, about 70\% of the source code (measured in bytes) is composed of Java files, 29\% are pages (HTML, JSP/JSF), and other file types account for 1\%. The systems have been in operation for over 13 years. They are used and customized by more than 30 other institutions.

The development teams of these three systems consist of 27 developers who use modern agile methodologies and practices. They are organized into \emph{software evolution teams} and \emph{software maintenance teams}. Software evolution teams work on the development of new features and architectural changes that have a significant impact on the system. Software maintenance teams are responsible for fixing bugs and making minor adjustments. Table~\ref{tab:systems_overview} shows the target teams and systems characterization.

\begin{table}[htbp]\centering
    \caption{Target Teams and Systems Characterization}
    \label{tab:systems_overview}
    \begin{tabular}{lrrrr}
        \hline
        System & Developers & Classes & Lines of code & Daily Requests \\ 
        \hline
        SIGAA & 14 & 8,156 & 1,323,196 & 1,208,326 \\
        SIPAC & 8 & 7,499 & 1,419,655 & 175,586 \\
        SIGRH & 5 & 4,405 & 778,588 & 92,877  \\
        \hline
        Total & 27 & 20,060 & 3,521,439 & 1,476,789  \\
        \hline
    \end{tabular}
\end{table}

\begin{figure}[htbp]
\centerline{
    \includegraphics[width=\columnwidth]{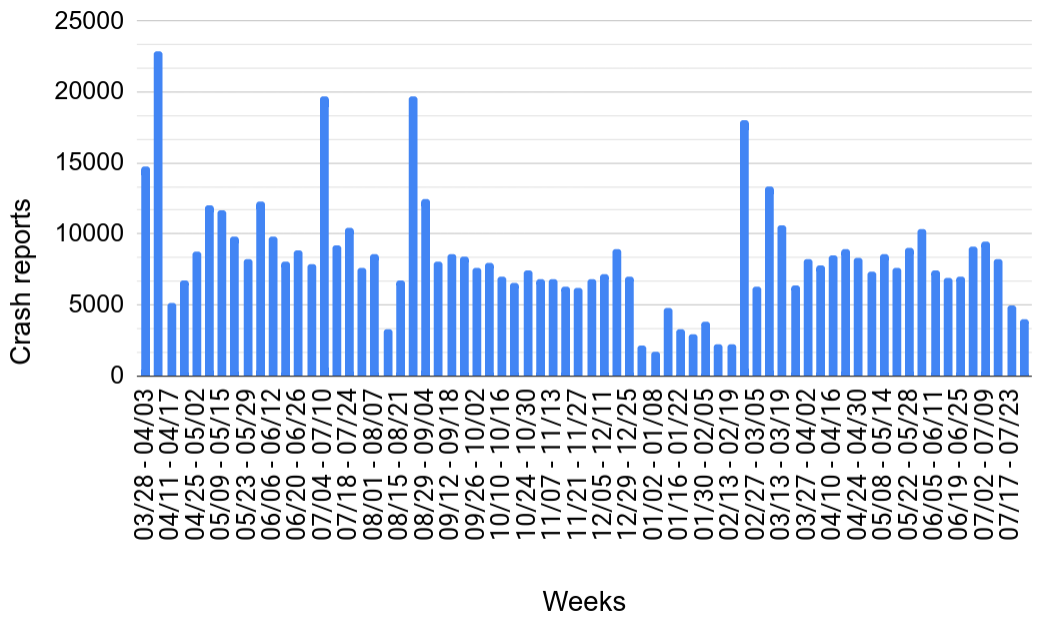}
}
\caption{SIGAA --- Amount of weekly crash reports}
\label{fig:sys1_crash_reports_vs_weeks}
\end{figure}

\begin{figure}[htbp]
\centerline{
    \includegraphics[width=\columnwidth]{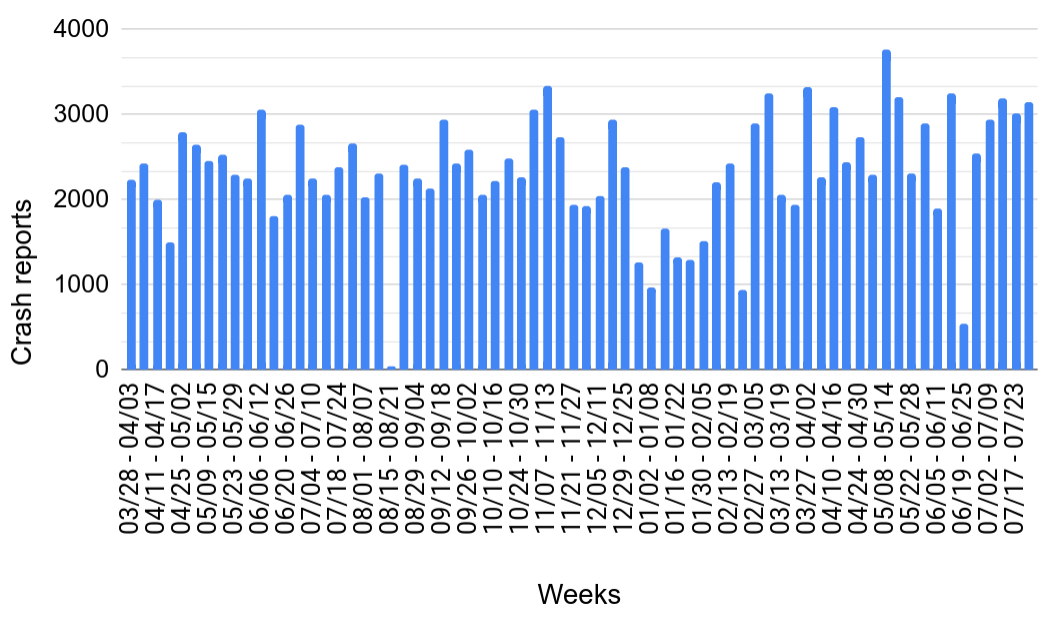}
}
\caption{SIPAC --- Amount of weekly crash reports}
\label{fig:sys2_crash_reports_vs_weeks}
\end{figure}

\begin{figure}[htbp]
\centerline{
    \includegraphics[width=\columnwidth]{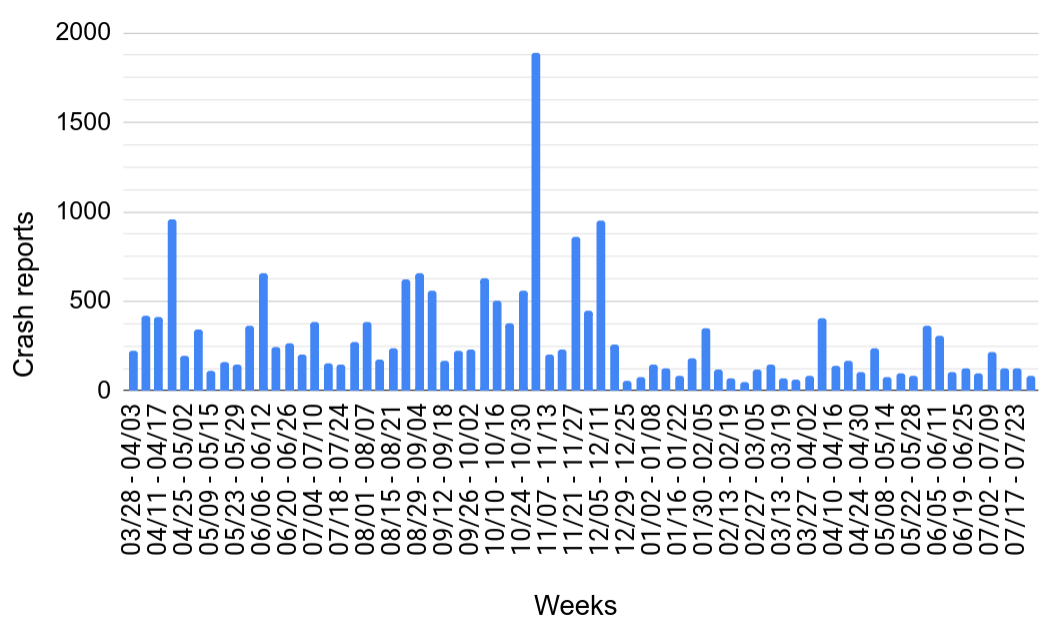}
}
\caption{SIGRH --- Amount of weekly crash reports}
\label{fig:sys3_crash_reports_vs_weeks}
\end{figure}

The first system (SIGAA) in Table~\ref{tab:systems_overview} is an Integrated Management System for Academic Activities. It is the most used and has the highest number of weekly crash reports, ranging from 1704 to 22895 crashes per week between March 2022 and August 2023, as shown in Fig. \ref{fig:sys1_crash_reports_vs_weeks}. SIPAC is an Integrated System for Asset Management, Administration, and Contracts. It presents the second-highest number of error reports. Fig.~\ref{fig:sys2_crash_reports_vs_weeks} shows that SIPAC has a number of crash reports close to zero in the week from August 15th to 21st of 2022. This was due to a maintenance effort on the machines and operating systems that resulted in the loss of logs. Finally, SIGRH is an Integrated Human Resource Management System. It has the lowest number of lines of code and daily requests and the lowest number of error logs, as shown in Fig. \ref{fig:sys3_crash_reports_vs_weeks}. The peak between October 31st and November 6th is due to several problems with the database infrastructure.


Regarding the crash report groups, SIGAA presented a greater number of groups (705), followed by SIPAC (442) and SIGRH (255), and also groups with more elements. 




\subsection{Study Procedures}\label{study_procedures}

To build a general understanding of the impact of the crash report clustering and suspicious source code ranking techniques in a real company, we implemented, customized, and applied an approach based on previous work~\cite{Wang2016, Wu, medeiros2020} for grouping crash reports and finding buggy code.

\begin{figure*}[htbp]
    \includegraphics[width=\textwidth]{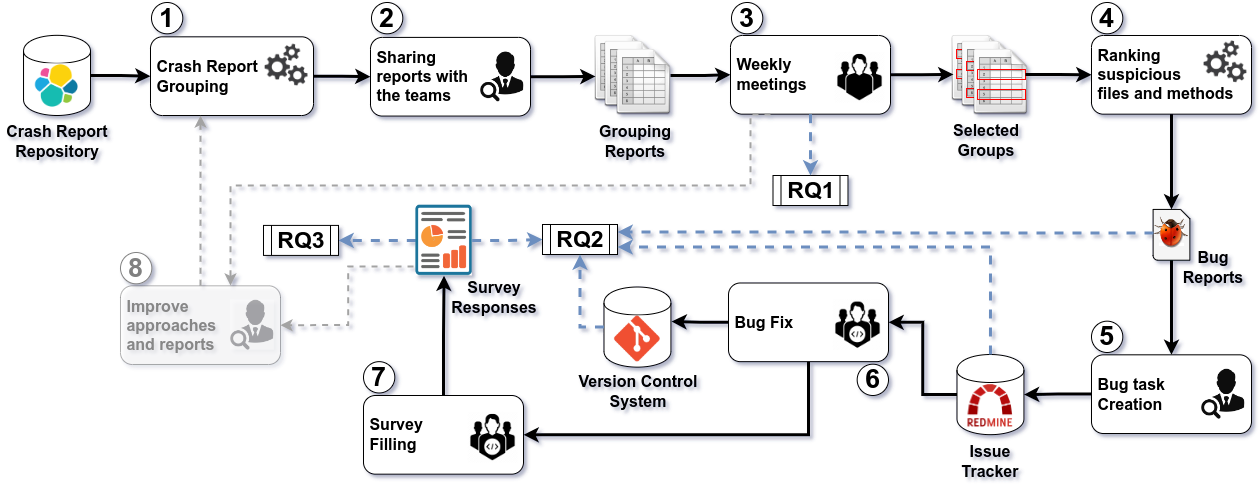}
    \caption{Study overview}
    \label{fig:fig_flow}
\end{figure*}

Fig. \ref{fig:fig_flow} gives an overview of the steps we conducted in our study. On a regular weekly basis, we first group existing crash reports from the investigated systems (Step 1 in Fig. \ref{fig:fig_flow}). After that, we expose the information about the crash report groups (Step 2 in Fig. \ref{fig:fig_flow}) to the project leaders to allow the prioritization of the ones to be examined in the current week (Step 3 in Fig. \ref{fig:fig_flow}). Then, we open bug issues related to the chosen crash report groups in the company-wide issue tracker system (Steps 4 and 5 in Fig. \ref{fig:fig_flow}).
Finally, we surveyed and interviewed the teams to collect information about the developers' perceptions (Step 7 in Fig. \ref{fig:fig_flow}). We processed crash reports and collected answers between April/2022 and August/2023. Next, we detail each of these steps.

\textit{{\bfseries Step 1 --- Crash report grouping}}. This step groups crash reports related to the same bug. 
We collect crash reports from the previous week (1 week before the meeting) in the production environment. The crash reports are then clustered as mentioned before (Section \ref{sec:crash_report_grouping}). As a result of the clustering approach, we get a set of crash report groups following the level 4 heuristic presented in Section \ref{sec:crash_report_grouping}. 

\textit{{\bfseries Step 2 --- Sharing crash report groups with the team}}. The purpose of this step is to allow developers to analyze and select the most important crash report groups. Every week, we share a spreadsheet that contains the following information for each group: identifier, first and last occurrence date, amount of crash reports, number of affected URIs, the total of affected users, and list of all system classes appearing in the stack traces. Based on that information, the development teams could understand the impact of a bug and which parts of the software are involved.

\textit{{\bfseries Step 3 --- Weekly meeting}}. Every week, we meet with the system’s project manager and technical leader to define the log groups for which we provide suspicious file rankings and generate the bug issues to be solved. During the meeting, we collect information about the experience of using the approach and discuss new directions on how we can improve it---such as including other relevant details in the bug report to facilitate the bug fix task. We also collect reasons for choosing each of the selected groups.

\textit{{\bfseries Step 4 --- Ranking suspicious files and methods}}. This step aims to facilitate the bug localization task, indicating files and methods that are more likely to have contributed to a crash. For each crash report group selected by the team, we mine the information of the files and methods present in the stack traces and use this information to rank these assets (files and methods) based on previous work~\cite{Wu, medeiros2020}. As detailed before (Section \ref{sec:ranking_suspicious_files}), we rank the files using three discriminative factors: Inverse Average Distance to crash Point (IAD), which considers that if a file appears closer to the crash point, it is more likely to cause the crash; Inverse Bucket Frequency (IBF) considers that if a file appears in stack traces caused by many different faults, it is less likely to be the cause of a specific fault; and File Frequency (FF), whose idea is that if a file often appears in stack traces caused by a particular defect, then it is likely to be the cause of this fault. 
We generate the ranking of the suspicious files using a combination of these three factors to produce a score for each file present in the stack traces of the group and sort them in descending order. For each file in the Top 5 suspicious list, we rank the suspicious methods based on how often they appear in stack traces.

\textit{{\bfseries Step 5 --- Bug tasks creation}}. In this step, we open tasks in the company's issue tracker system with additional information to help fix the bug. We create an issue with the Top 5 suspicious files and methods produced in previous steps for each selected crash report group. In addition, we include the first and last date of the error log occurrence, the number of logs in the group, the Top 5 affected URIs with respective Top 5 affected users, and how many times they were involved. We also add stack trace samples, crash report IDs (the unique identifier for a crash report on the database), and the user's sign-in session identifiers. This additional information was mined from the crash report database. We also include instructions for the developers who are going to fix the bug, asking them (a) not to combine code refactoring with bug fixes in the same commit, (b) to reference the bug-fixing task in the commit, and (c) to fill out the survey after closing the issue. 

\textit{{\bfseries Step 6 --- Bug fixing}}. At this stage, teams prioritize the opened issues and distribute them to developers who try to identify and fix the bugs.

\textit{{\bfseries Step 7 --- Survey filling}}. This step aims to collect the developers' perceptions about the application of the approach and the provided information in the bug-fixing tasks. After identifying and fixing the issue, the developers accessed an online survey we implemented and filled it out. The survey questions whether the bug was present in the suggested classes and methods, whether the ranks of files and methods helped resolve the bug, and whether each provided information was helpful in the fix task. Developers can also submit criticisms and suggestions about the approach and data provided in the tasks. Table \ref{tab:survey_questions} shows the survey questions.

\begin{table}[htbp]\centering
    \caption{Survey questions}
    \label{tab:survey_questions}
    \begin{tabular}{p{0.04\textwidth}p{0.4\textwidth}}
        \hline
        Number & Question \\ 
        \hline
        SQ1 & What task number did you solve? \\
        SQ2.1.1 & Did the list of suspicious files help to fix the bug? \\
        SQ2.1.2 & Leave your comments (criticism, suggestions, etc.) about the suspicious files suggested by the approach. \\
        SQ2.2.1 & Did the list of suspicious methods help to fix the bug? \\
        SQ2.2.2 & Leave your comments (criticism, suggestions, etc.) about the suspicious methods suggested by the approach \\
        SQ3.1 & Did the list of examples of affected URIs help to fix the bug? \\
        SQ3.2 & Did the list of examples of affected users help to fix the bug? \\
        SQ3.3 & Did the stack trace(s) example(s) help to fix the bug? \\
        SQ3.4 & Did the list of examples of related log IDs help to fix the bug? \\
        SQ3.5 & Leave your comments (criticisms, suggestions, etc.) about the lists of URIs, affected users, stack traces, and related log IDs. \\
        SQ3.6 & Is there any additional information you believe will help with bug fixes? Or any other improvement suggestions for the tool? Comment below. \\
        \hline
    \end{tabular}
\end{table}

\section{Results and Discussion} \label{sec:results_and_discussion}

This section reports our results regarding the criteria developers use to prioritize crash report groups and how useful the provided crash report information is to solve associated bugs. We present the results of the research questions we introduced before.

Overall, we have opened 131 bug issues, of which 86 have been closed, as detailed in Table \ref{tab:tasks_overview}. Among the 86 (100\%) closed bug issues, 50 (58\%) have at least one commit referencing it that modified any Java file. We analyzed the 36 (42\%) tasks without committed Java files and identified that 12 had associated commits but modified other types of files (e.g., Jasper, jrxml, JSP). The other 24 tasks did not have any related commits, whereas for 23 of them, the teams could not reproduce the bug, and for a specific one, the error was caused by a failure in the database.

\begin{table}[htbp]\centering
    \caption{Bug fix tasks overview}
    \label{tab:tasks_overview}
    \begin{tabular}{lrrrrr}
        \hline
         Team & Opened & Closed & With committed Java File \\ 
        \hline
        Team1 (SIGAA) & 81 & 37 & 15 \\
        Team2 (SIPAC) & 11 & 11 & 1 \\
        Team3 (SIGRH) & 39 & 38 & 34 \\
        \hline
        Total & 131 & 86 & 50 \\
        \hline
    \end{tabular}
\end{table}

\subsection{\rqa}

In our interactions with the development team, we clustered crash reports and suggested suspicious files to the groups with the highest number of crash occurrences because this is a common way that software developers prioritize crash reports \cite{an2015challenges}. We based our recommendations on the assumption that larger groups affect more users and should be resolved first \cite{kim2011crashes}. Surprisingly, we observed that the larger groups were not always chosen by the development team to be analyzed first. We investigated RQ1 to understand the reasons for these choices.

We found several situations where the project manager and the technical leader preferred to first analyze and select groups with a few crash reports instead of other larger groups. For example, they chose groups with 37 and 14 crash reports instead of a group with more than 5000 crash reports in a specific week. During the weekly meetings, we asked them why they chose each group and why some larger groups were almost ignored. 
We identified five main motivations used by the teams to prioritize a group over another: 

\begin{itemize}
\item {\bfseries Critical requirements:} high-priority features for the business in that specific week or the following weeks;
\item {\bfseries Module modified recently:} modules in which some parts of source code were changed in the previous weeks;
\item {\bfseries Feature used intensively:} feature with a large volume of use in the analyzed period or that will be used intensively in the following weeks;
\item {\bfseries Number of error logs:} the amount of crash reports in the crash report group;
\item {\bfseries Complexity for reproducing the crash:} estimated time to reproduce and fix a bug.
\end{itemize}


Table \ref{tab:motivations_for_group_choices} shows the percentage that project leaders mentioned each motivation. It is worth noting that more than one reason can be used to choose groups. Overall, the participants prioritized crash report groups associated with system requirements that were more critical during that week (48.85\%). Next, they selected groups with the highest number of errors (29.77\%). Then, they mentioned the complexity of reproducing the bug (16.03\%) in the test environment or features intensively used by system users (15.27\%). Finally, with fewer mentions, they selected groups whose features had been modified (12.98\%) during that week and other factors (6.11\%). Technical leaders commented about other factors used to prioritize crash report groups, such as the expertise level of the developers available at the current week to be assigned to fix bug tasks and the estimated time to fix a bug. 

\begin{table*}[htbp]\centering
    \caption{Motivations for group choices}
    \label{tab:motivations_for_group_choices}
    \begin{tabular}{lrrrrrr}
        \hline
         &  & Module & Feature & Number & Complexity &\\ 
         & Critical & modified & used & of error & of reproducing & \\ 
         Team & requirement & recently & intensively & logs & the crash & Others \\ 
        \hline
        Team1 & 67.90\% & 19.75\% & 12.35\% & 13.58\% & 00.00\% & 08.64\% \\
        Team2 & 45.45\% & 00.00\% & 36.36\% & 36.36\% & 00.00\% & 00.00\% \\
        Team3 & 10.26\% & 02.56\% & 15.38\% & 61.54\% & 53.85\% & 00.00\% \\
        \hline
        Total & 48.85\% & 12.98\% & 15.27\% & 29.77\% & 16.03\% & 06.11\% \\
        \hline
    \end{tabular}
\end{table*}


Analyzing each team individually, we can see they use very distinct strategies. Team1 prioritized groups associated with more critical requirements (67.90\%) and features modified recently (19.75\%). At the same time, Team2 chose groups based on critical requirements (45,45\%), usage intensity (36.36\%), and group size (35.36\%). Team3 prioritized by group size (61,54\%) and easiness of bug reproduction (53.85\%).

We also investigated with the participants the reasons for not prioritizing some groups with the highest number of crash reports. The teams decided to postpone most of those groups because they are associated with critical architectural changes. These architectural changes are being addressed by a different senior team, which is proposing the migration of the current reference architecture implementation of the system to other modern technologies and frameworks. Therefore, these groups were not chosen to be investigated because the participants already knew their root causes, and the bugs associated with those large groups do not affect the successful execution of the system functionalities. 

\vspace{2mm}

\noindent\fbox{\begin{minipage}{0.97\columnwidth}
\textbf{Answer to RQ1}: Teams use multiple criteria to prioritize the resolution of crash report groups. Factors with high priority include: critical requirements, modules recently modified, intensively-used features, and the number of error logs. Other factors appear to have low priority (in other words, they made developers not choose crash report groups related to them), such as bug fixing related to architectural changes, the complexity of reproducing the crash, and the high estimated time to fix a bug.
\end{minipage}}

\subsection{\rqb}

To answer this question, we first checked if the changed files and methods were present in the list of suspicious files and methods. We accomplished this in a similar way that we used in our previous study \cite{medeiros2020}. After each bug-fixing task was completed, we collected information about committed files that fix the bug in the issue tracker system (Redmine), and we compared them with the suspicious files provided by our approach. In collaboration with the development teams, we extracted the names of the methods that have been changed in the bug-fixing commits by mining this information from the issue tracker system and Git code repository.

An average of 1.4 Java files were committed per task. For 82\% of the tasks, only one file was changed in their associated commits, two files were for 12\%, and three or more files were modified for 6\%. The average of methods committed per task was 2.24, where a single method was modified for 64\% of them, two methods for 18\%, and three or more for 18\%.

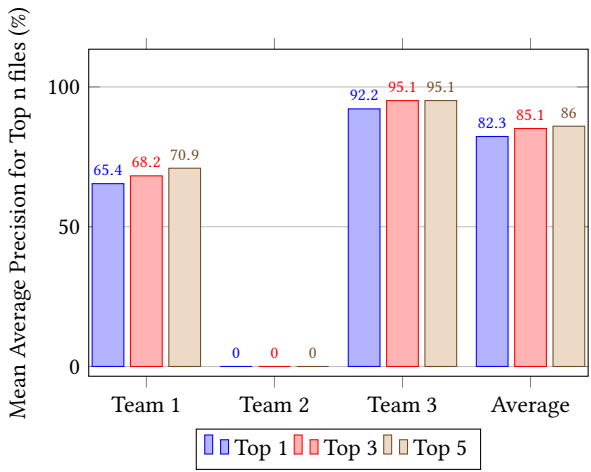
\begin{figure}[htbp]\centering
\pgfplotsset{
    ybar={2.5pt},
    xtick=data,
    symbolic x coords={Team 1, Team 2, Team 3, Average},
    width=0.97*\linewidth,
    height=0.7*\linewidth,,
    legend style={at={(0.5,-0.16)}, anchor=north, legend columns=-1},
    nodes near coords={
        \pgfmathprintnumber[precision=1]{\pgfplotspointmeta}
    },
    ymin=10, 
    ymax=100,
    enlargelimits=0.15,
    ymajorgrids=true,
    /pgf/bar width=12pt,
    every node near coord/.append style={font=\scriptsize}
}
\begin{tabular}{l}
    \begin{tikzpicture}
        \begin{axis}[
            ylabel=Mean Average Precision for Top n files (\%),
        ]
            \addplot coordinates {(Team 1, 65.40) (Team 2, 0.0) (Team 3, 92.15) (Average, 82.28) };
            \addplot coordinates {(Team 1, 68.20) (Team 2, 0.0) (Team 3, 95.12) (Average, 85.14) };
            \addplot coordinates {(Team 1, 70.93) (Team 2, 0.0) (Team 3, 95.12) (Average, 85.96) };
            \legend{Top 1, Top 3, Top 5}
        \end{axis}
    \end{tikzpicture}
\end{tabular}%
\caption{Mean Average Precision for Top N Files}
\label{fig:mean_map}
\end{figure}

\begin{figure}[htbp]\centering
\pgfplotsset{
    ybar={2.5pt},
    xtick=data,
    symbolic x coords={Team 1, Team 2, Team 3, Average},
    width=0.97*\linewidth,
    height=0.7*\linewidth,,
    legend style={at={(0.5,-0.16)}, anchor=north, legend columns=-1},
    nodes near coords={
        \pgfmathprintnumber[precision=1]{\pgfplotspointmeta}
    },
    ymin=10, 
    ymax=100,
    enlargelimits=0.15,
    ymajorgrids=true,
    /pgf/bar width=12pt,
    every node near coord/.append style={font=\scriptsize}
}
\begin{tabular}{l}
    \begin{tikzpicture}
        \begin{axis}[
            ylabel=Recall@N (\%),
        ]
            \addplot coordinates {(Team 1, 86.67) (Team 2, 0.0) (Team 3, 94.12) (Average, 90.00) };
            \addplot coordinates {(Team 1, 86.67) (Team 2, 0.0) (Team 3, 100.00) (Average, 94.00) };
            \addplot coordinates {(Team 1, 93.33) (Team 2, 0.0) (Team 3, 100.00) (Average, 96.00) };
            \legend{Top 1, Top 3, Top 5}
        \end{axis}
    \end{tikzpicture}
\end{tabular}%
\caption{Recall@N}
\label{fig:mean_recall}
\end{figure}
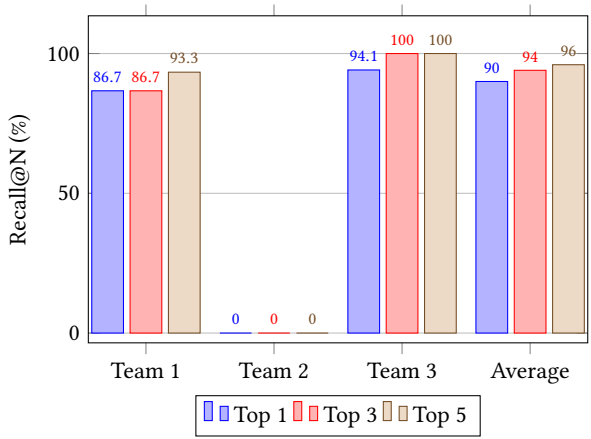

We measured the performance of our approach using the Recall@N and MAP metrics in a similar way that we used in our previous study \cite{medeiros2020}. Figs. \ref{fig:mean_map} and \ref{fig:mean_recall} show the results, considering the ranked list of Top with 1, 3, and 5 suspicious files. We can see that MAP, on average (Fig. \ref{fig:mean_map}), ranges between 82\% and 86\%, while recall (Fig. \ref{fig:mean_recall}) ranges between 90\% and 96\%. These values are higher than those obtained in our previous retrospective study \cite{medeiros2020}, except for the Team 2, where only one closed task had a committed Java file, and no suspicious files were modified in the commit.

We also observed that, for 94\% of the tasks, at least one method modified in the commits was present in the suspect method lists provided by our approach. The percentage was 86.67\% for SIGAA tasks and 100\% for SIGRH tasks. Our approach usually suggests one or two methods for each suspicious file because only a few methods appear in the stack traces of a group of logs. We found that a single method appeared for 46\% of the suspicious files suggested by the approach, two for 21\%, and three or more methods appeared for 33\%. A Java class in the analyzed projects commonly has over a dozen methods, so suggesting one or two methods from each suspicious file can help decrease the effort to fix the bug.


Next, we analyzed the developers' opinions about our approach, assessing the responses to the survey questions SQ2.1.1 and SQ2.2.1 (see Table \ref{tab:survey_questions}). We obtained feedback (answers in our survey form) for 67 of the 86 closed tasks. It is important to highlight that developers can complete the survey for all closed tasks. Usually, when a task is closed without fixing the bug, they fill out the form mentioning that the provided information does not help to fix the bug. They are also not required to answer the questionnaire, so we only have 67 answers, not 86.

\begin{figure}[htbp]\centering
\pgfplotsset{
    ybar={2.5pt},
    xtick=data,
    symbolic x coords={Team 1, Team 2, Team 3, Average},
    width=0.97*\linewidth,
    height=0.7*\linewidth,,
    legend style={at={(0.5,-0.16)}, anchor=north, legend columns=-1},
    nodes near coords={
        \pgfmathprintnumber[precision=1]{\pgfplotspointmeta}
    },
    ymin=10, 
    ymax=90,
    enlargelimits=0.15,
    ymajorgrids=true,
    /pgf/bar width=12pt,
    every node near coord/.append style={font=\scriptsize}
}
\begin{tabular}{l}
    \begin{tikzpicture}
        \begin{axis}[
            ylabel=Helped (\%),
        ]
            \addplot coordinates {(Team 1, 70.00) (Team 2, 0.0) (Team 3, 74.19) (Average, 65.67) };
            \addplot coordinates {(Team 1, 60.00) (Team 2, 0.0) (Team 3, 64.52) (Average, 56.72) };
            \legend{Files, Methods}
        \end{axis}
    \end{tikzpicture}
\end{tabular}%
\caption{Closed bug fix tasks: Do suspicious files and methods help to fix the bug?}
\label{fig:do_files_and_methodos_help_fix_the_bug}
\end{figure}
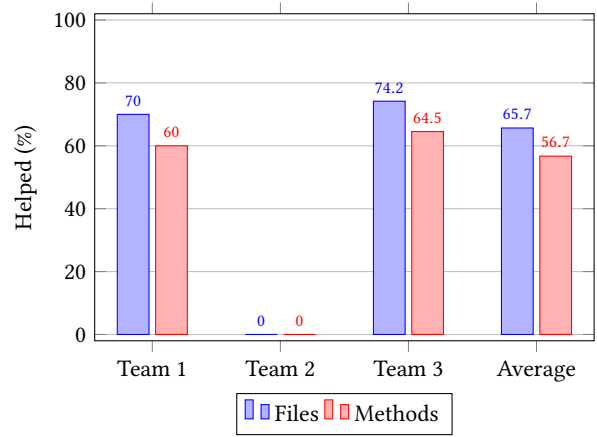

In total, 18 developers of different experience levels answered the questionnaire. Table \ref{tab:developers_experience} shows the percentage of responses to the questionnaire by developers' experience level. We can see that 43\% of the answers were from senior developers, 14\% from middle, and 43\% from junior developers. Fig. \ref{fig:do_files_and_methodos_help_fix_the_bug} shows that, on average, developers considered the \emph{file} and \emph{method ranking} information useful in 65.7\% and 56.7\%  of closed tasks, respectively. Our study participants could not identify and reproduce the errors reported in 28\% (24 of 86) of the closed tasks. These data reflect the difficulty of reproducing bugs in the testing environment~\cite{kilgore1997re, Cukier2013, erfani2014works, rahman2020some}.

\begin{table}[htbp]\centering
    \caption{Responding Developers' Profile by Experience Level}
    \label{tab:developers_experience}
    \begin{tabular}{lrrrrr}
        \hline
         Team & Junior & Middle & Senior \\ 
        \hline
        Team1 & 70\% & 17\% & 13\% \\
        Team2 & 33\% & 67\% & 0\% \\
        Team3 & 23\% & 0\% & 77\% \\
        \hline
        Total & 43\% & 14\% & 43\% \\
        \hline
    \end{tabular}
\end{table}

\begin{figure}[tbp]\centering
\pgfplotsset{
    ybar={2.5pt},
    xtick=data,
    symbolic x coords={Team 1, Team 2, Team 3, Average},
    width=0.97*\linewidth,
    height=0.7*\linewidth,,
    legend style={at={(0.5,-0.16)}, anchor=north, legend columns=-1},
    nodes near coords={
        \pgfmathprintnumber[precision=1]{\pgfplotspointmeta}
    },
    ymin=10, 
    ymax=90,
    enlargelimits=0.15,
    ymajorgrids=true,
    /pgf/bar width=12pt,
    every node near coord/.append style={font=\scriptsize}
}
\begin{tabular}{l}
    \begin{tikzpicture}
        \begin{axis}[
            ylabel=Helped (\%),
        ]
            \addplot coordinates {(Team 1, 91.67) (Team 2, 0.0) (Team 3, 75.00) (Average, 78.05) };
            \addplot coordinates {(Team 1, 75.00) (Team 2, 0.0) (Team 3, 64.29) (Average, 65.85) };
            \legend{Files, Methods}
        \end{axis}
    \end{tikzpicture}
\end{tabular}%
\caption{Bug fix tasks with committed Java Files: Do suspicious files and methods help to fix the bug?}
\label{fig:do_files_and_methodos_help_fix_the_bug_with_commited_java_files}
\end{figure}
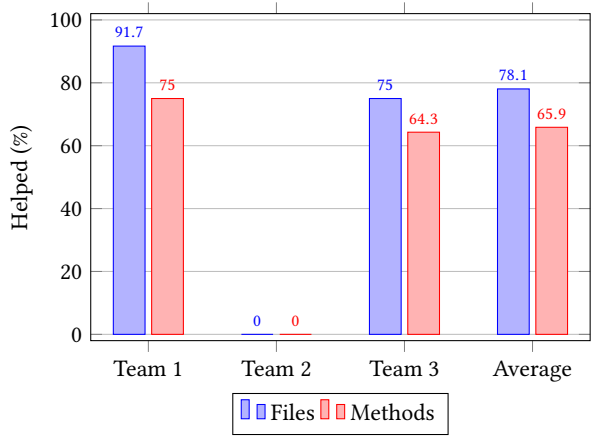

We also measured developers' perceptions considering only the 50 tasks with committed Java files to check for a difference when the tasks were closed by bug fixing. The results were better than those obtained in Fig. \ref{fig:do_files_and_methodos_help_fix_the_bug}. Fig. \ref{fig:do_files_and_methodos_help_fix_the_bug_with_commited_java_files} shows the percentages by development teams and the overall average considering answers related to the tasks with bug-fix commits and modified Java files. The ranking of suspicious files helped developers fix bug reports for 78.1\% of the tasks, and ranking suspicious methods helped for 65.9\% of the tasks. As such, we believe that improving the reproduction of errors in the testing environment can help to improve the approach's effectiveness. The developers' perception of the usefulness of the list of suspicious files and methods improves recall and MAP numbers. Altogether, these findings suggest that our approach is feasible and can help developers in bug-fixing tasks.

For the survey question about suspicious file ranking (SQ2.1.2), we received a comment that the bug had already been fixed in another task, but it was in the suspicious file rank. In another answer, a participant wrote that it was only possible to resolve the bug after consulting information unavailable in the task. Another developer said she/he liked the rank because it increased the conviction of which code to fix. Other answers mentioned that the bug could not be reproduced. Even so, for some cases of non-reproduction of the error, some developers wrote that it was possible to make interventions to mitigate the crash. These answers indicate that the approach can be helpful in the daily life of a software company and that a possible way to improve it is to provide additional information when opening bug-fixing tasks.

By cross-referencing the information in Table \ref{tab:developers_experience} and the data in Fig. \ref{fig:do_files_and_methodos_help_fix_the_bug}, we cannot find any difference in developers' perceptions depending on their experience levels. However, one of the team leaders reported that ranking suspicious files and methods and the additional information provided by our approach especially helps beginner developers.

In the survey question about suspicious methods (SQ2.2.2), a developer commented about the list of suspicious methods. She/he wrote that the root cause of the bug is not always in the method where the exception was raised. Sometimes, the crash occurs in a method, but the root cause is in another method or class. This developer’s comment is related to the fact that our approach is based on stack traces, and the methods that contain crashes do not always reside in the crash stack traces, as described in \cite{an2015challenges, Wu}. We discussed this limitation with the project leaders at the beginning of our study. Other answers commented that including the lines of the suspect methods where the crash occurred would be interesting, although we have added examples of stack traces in the tasks. One developer commented that the methods presented as suspect helped to identify the source of the problem, although this source is not coming from the methods.

We analyzed whether any crash reports from the error group addressed by the bug tasks continued appearing after their closure (in this case, the bug was only closed when they were fixed). Table \ref{tab:groups_that_continued_with_logs} shows that error logs were no longer present in the crash report database for most tasks (68\%). These results reinforce that using the approach on a day-to-day basis helps developers to fix the bug. However, for some error logs (32\%), we could observe that they re-appear in the production environment. These cases may occur due to partial bug fixes or bug reintroductions.

\begin{table}[htbp]\centering
    \caption{Bug fix tasks whose groups continued to manifest error logs after the task was completed (closed)}
    \label{tab:groups_that_continued_with_logs}
    \begin{tabular}{l|rr}
        \hline
        Error log group continue to appear? & No & Yes \\ 
        \hline
        Team1 & 9 (60.00\%) & 6 (40.00\%) \\
        Team2 & 0 (0.00\%) & 1 (100.00\%) \\
        Team3 & 25 (73.53\%) & 9 (26.47\%) \\
        \hline
        Total & 34 (68.00\%) & 16 (32.00\%)  \\
        \hline
    \end{tabular}
\end{table}

\noindent\fbox{\begin{minipage}{0.97\columnwidth}
\textbf{Answer to RQ2}: The approach correctly suggested at least one buggy file among the top-5 96\% of the time --- with mean average precision above 86\% --- and at least one buggy method 94\% of the time. The developers were surveyed and confirmed the effectiveness of using the approach. They also reported that ranking suspicious files and methods helped fix the bug, even when it was not present in any suggested files or methods.
\end{minipage}}

\subsection{\rqc}

During the study, the development team asked us to add other information they consider relevant (e.g., affected users, crash report IDs)---in addition to the data provided by our approach (suspicious files and methods, and stack trace samples). The requested information was available in the crash reports. We then improved our approach by processing and including this information when opening the bug-fixing tasks. In this way, we addressed our third research question (RQ3), which aims to understand what additional information can help developers fix bugs. 

We currently provide four types of additional information for each crash report group: (i) URIs affected with the respective number of occurrences; (ii) top 5 affected users in each affected URI with the respective number of occurrences; (iii) sample of a stack trace; and (iv) examples of crash report IDs. We included a multiple-choice question (yes or no) in the survey for each type of additional information provided (SQ3.1, SQ3.2, SQ3.3, and SQ3.4 of Table~\ref{tab:survey_questions}). 

\begin{figure}[htbp]\centering
\pgfplotsset{
    ybar={2.5pt},
    xtick=data,
    symbolic x coords={STs, URIs, CR IDs, Users},
    width=0.97*\linewidth,
    height=0.7*\linewidth,
    legend style={at={(0.5,-0.16)}, anchor=north, legend columns=-1},
    nodes near coords={
        \pgfmathprintnumber[precision=0]{\pgfplotspointmeta}
    },
    ymin=1,
    ymax=100,
    enlargelimits=0.15,
    ymajorgrids=true,
    /pgf/bar width=9pt
}
\begin{tabular}{l}
    \begin{tikzpicture}
        \begin{axis}[
            ylabel=\%,
        ]
            \addplot coordinates {(STs, 70.00) (CR IDs, 60.00) (URIs, 70.00) (Users, 60.00)};
            \addplot coordinates {(STs, 16.67) (CR IDs, 16.67) (URIs, 16.67) (Users, 33.33)};
            \addplot coordinates {(STs, 83.87) (CR IDs, 83.87) (URIs, 35.48) (Users, 29.03)};
            \addplot coordinates {(STs, 71.64) (CR IDs, 41.79) (URIs, 49.25) (Users, 43.28)};
            \legend{Team 1, Team 2, Team 3, Average}
        \end{axis}
    \end{tikzpicture}
\end{tabular}%
\caption{Closed bug fix tasks: Additional information helps fix bug}
\label{fig:additional_information_help_fix_bug}
\end{figure}
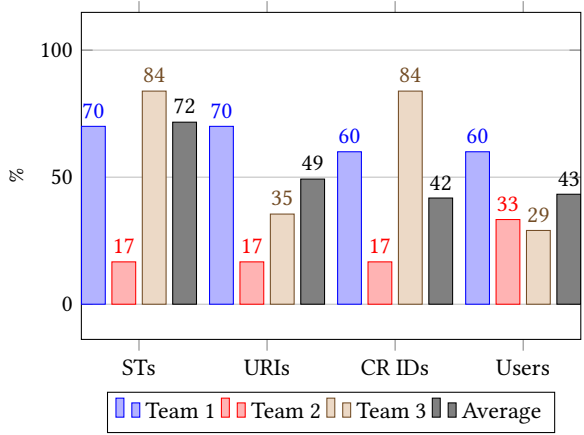

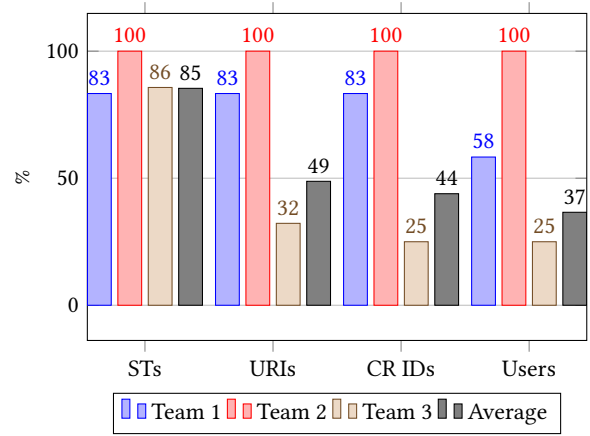
\begin{figure}[htbp]\centering
\pgfplotsset{
    ybar={2.5pt},
    xtick=data,
    symbolic x coords={STs, URIs, CR IDs, Users},
    width=0.97*\linewidth,
    height=0.7*\linewidth,
    legend style={at={(0.5,-0.16)}, anchor=north, legend columns=-1},
    nodes near coords={
        \pgfmathprintnumber[precision=0]{\pgfplotspointmeta}
    },
    ymin=1,
    ymax=100,
    enlargelimits=0.15,
    ymajorgrids=true,
    /pgf/bar width=9pt
}
\begin{tabular}{l}
    \begin{tikzpicture}
        \begin{axis}[
            ylabel=\%,
        ]
            \addplot coordinates {(STs, 83.33) (CR IDs, 83.33) (URIs, 83.33) (Users, 58.33)};
            \addplot coordinates {(STs, 100.00) (CR IDs, 100.00) (URIs, 100.00) (Users, 100.00)};
            \addplot coordinates {(STs, 85.71) (CR IDs, 25.00) (URIs, 32.24) (Users, 25.00)};
            \addplot coordinates {(STs, 85.37) (CR IDs, 43.90) (URIs, 48.78) (Users, 36.59)};
            \legend{Team 1, Team 2, Team 3, Average}
        \end{axis}
    \end{tikzpicture}
\end{tabular}%
\caption{Bug fix tasks with committed Java Files: Additional information helps fix bug}
\label{fig:additional_information_help_fix_bug_with_commited_java_files}
\end{figure}

In a similar way to what we did when checking whether the files and methods helped fix the bug (Fig. \ref{fig:do_files_and_methodos_help_fix_the_bug} and Fig. \ref{fig:do_files_and_methodos_help_fix_the_bug_with_commited_java_files}), we analyzed the developers' perceptions considering all closed issues (Fig. \ref{fig:additional_information_help_fix_bug}) and considering only the tasks with Java committed files (Fig. \ref{fig:additional_information_help_fix_bug_with_commited_java_files}). In both cases, stack traces (STs) and URIs helped the most. The most significant difference regarding the averages is the stack traces, which was 72\% in Fig. \ref{fig:additional_information_help_fix_bug} and 85\% in Fig. \ref{fig:additional_information_help_fix_bug_with_commited_java_files}. Affected URIs (URIs) helped in 49\% of the issues. Examples of crash report IDs (CR Ids) were helpful between 42\% and 44\% of the issues, and the list of affected users (Users) helped fix the bug between 37\% and 43\% of the issues. Overall, the additional information in the tasks helped to fix the bug, but it is worth noting that the examples of stack traces helped more than the list of suspicious files and methods.

We also included two survey questions (SQ3.5 and SQ3.6) to collect comments, criticisms, and suggestions about the additional information. Regarding the SQ3.5 (Table~\ref{tab:survey_questions}) question, we obtained responses that the data provided was interesting, detailed, and direct. On the other hand, another developer commented that she/he felt confused as there were too many stack traces and no example logins. In this case, the failure occurred on screens of the web system where authentication is not required, and we identified several stack traces related to the same error. 
In another answer, it was reported that while only the stack trace helped to fix the bug, the rest of the information is essential for further analysis. Finally, a developer commented that the URLs and crash report IDs were useful in identifying the system functionality that has the error, but they were not enough to find its cause (intermittent error).

The answers to question SQ3.6 (Table~\ref{tab:survey_questions}) are predominantly related to error reproducibility. Some developers reported that they could not simulate the error even after talking to users and re-executing the same steps. They believe that the errors occurred due to some momentary instability of the system. We also had the following suggestions for improvements: (i) include logins of affected users; (ii) decrease the amount of stack traces included in tasks; and (iii) add the steps performed in the system use case for the error to occur, in addition to the tool being able to reproduce the reported error.

We already used the developers' feedback to add more details to our approach, such as the affected users sorted by the number of times each of them were involved, to help fix bugs. We also limited the number of URIs, user logins, and stack traces we included in the tasks to reduce the volume of information and minimize confusion, especially for beginner developers.

\vspace{2mm}
\noindent\fbox{\begin{minipage}{0.97\columnwidth}
\textbf{Answer to RQ3}: Examples of stack traces help with most tasks. Samples of affected URIs, crash report IDs, and user logins helped in a minor proportion.
\end{minipage}}



\section{Implications and Discussions} \label{sec:lessons_learned}








\subsection{Implications to Practitioners}

Our results of applying the crash report grouping approach and ranking of code suspected of containing the bug (Section 4.2, RQ2) were better than those obtained in our retrospective study for the same target systems \cite{medeiros2020}. They show that the list of suspicious files and methods can strongly contribute to fixing the bugs related to the logged crashed reports in large web-based systems. This encourages the adoption of these techniques in the daily life of development teams. Indeed, the teams in which we applied the approach have been using it weekly since March 2022. 

Our experience has shown that the approach usage should consider summarizing the log group data for the developers, avoiding unnecessary information, and including relevant information to reproduce the bug, such as samples of stack traces, affected URIs, crash report IDs, and user logins. This information facilitates the bug's reproduction, discovery, and resolution. 

Regarding the usage of our approach by developers and project leaders, we have initially implemented a web system that allows them to search and navigate along the different crash report log group data, getting information such as identifier, first and last occurrence date, amount of crash
reports, number of affected URIs, the total of affected users, and list of all system classes appearing in the stack traces. However, the project leaders who are responsible for indicating which log group we should use to create new bug tasks (see Fig. \ref{fig:fig_flow}), prefer that we create simple spreadsheets with summarized information of the log groups that happened in the week under analysis. They then use these spreadsheets to select the log groups to be solved as a new bug-fix issue during that week. The information available in such spreadsheets has been refined based on their feedback.  

The developers assigned to correct bug-fix issues created by the project leaders access all the summarized information related to a chosen log group directly in the Redmine\footnote{https://www.redmine.org/} issue track system adopted by the company. With that information, the developers can also use the Kibana\footnote{https://www.elastic.co/kibana} software to visualize the Elastic Search\footnote{https://www.elastic.co/elasticsearch} logs and investigate the bugs' root causes. They analyze, for example, the executed system functionalities operations of users during their interaction with the system to better understand the context of the crash report occurrence. Elastic database stores crash reports and user interaction logs (all HTTP requests with their parameters). The developers usually look for the crash report by ID and observe the associated stack trace. When they cannot identify the cause, they try to reproduce the functionalities executed by the user that crashed the system. In this case, they analyze the interaction logs associated with the user's session. It shows that different tools must be available to be used together with the summarized information of the log groups to facilitate bug reproduction.  

\subsection{Implications to Researchers}

For researchers interested in this topic, a potential extension of this work could be to apply the approach with teams from other companies and/or using other technologies and programming languages. This could contribute to confirming the obtained results and provide insights into different development contexts.

Additionally, the approach could be enhanced to include additional information that assists the teams with prioritizing tasks of error log groups. For example, to include criticality information of the impacted functionalities from a business perspective, to show whether any code snippets in the stack traces were recently modified, or to indicate whether any functionality in the log group is being intensively utilized during the analyzed period. These details could help developers to prioritize crash reports more effectively.

The study also identified that reproducing errors can be challenging for development teams~\cite{erfani2014works, rahman2020some}. Therefore, an area of future research could be to understand the reasons for this difficulty and provide additional information to assist developers in this task. Applying techniques for automated test generation~\cite{shamshiri2015automatically} to the list of suspicious files and methods suggested by the approach could provide more insights to developers to assist with debugging and reproducing errors and reduce the time required to resolve issues.

\section{Threats to Validity} \label{sec:threats_to_validity}

\textit{{\bfseries Construct validity threats.}}
We may have made measurement errors since we obtained data from stack traces to group them using regular expressions in a step of our study. To mitigate such threats, we have carefully codified and tested our implementation. Bug-fixing commits can also have changes associated with code refactoring. In our study, 82\% of bug-fixing commits have modified only one file, and 64\% have modified only one method. It shows a high probability that most files and methods are associated with bugs, not refactoring. 

\textit{{\bfseries Internal Validity Threats.}}
We extracted information from stack traces stored in Elastic Search and obtained changed files for bug fixes via webhooks\footnote{https://developer.github.com/webhooks/} from Redmine. It is important to note that this information may be incomplete, and their data quality could impact the accuracy of fault localization.

\textit{{\bfseries External Validity Threats.}} We analyzed issues and crash reports for three systems and development teams from a single company, all implemented using Java technologies. These systems may reflect the characteristics commonly found in a specific software category, such as Java web-based information systems. To extend the applicability of our findings, further research should be conducted with a broader range of systems.

\section{Related Work} \label{sec:related_work}

Over the last years, several research work has developed, improved, and applied crash report clustering approaches ~\cite{podgurski2003automated,khomh2011entropy,kim2011crashes,dang2012rebucket,Wang2013,Wang2016} to facilitate bug prioritization and localization. Also, other research work has developed techniques to rank buggy files and methods and find the root cause of the crashes~\cite{ball2003symptom,jones2002visualization,jones2005empirical,nessa2008software,schroter2010stack,wong2014boosting,gu2019does,Wu,wu2018changelocator}. Most of the existing work used data from open source systems, such as Firefox, Thunderbird, and Eclipse, because their crash report data is open to the public. 

Jarman et al. \cite{jarman2021legion} applied the Legion (an extension of BugLocator) in an industrial context at Adobe Analytics and reported a substantial improvement over BugLocator. In a retrospective study, they found that 76\% of the time, at least one buggy file ranked in the Top 10 recommendations. They also conducted a pilot study with three developers, asking the developers who fixed the issues to evaluate the effectiveness of the prediction. Their results show junior developers will likely benefit from automated bug localization tools. Li and colleagues \cite{li2022empirical} conducted a study applying six state-of-the-art IRBL techniques to 10 Huawei projects and compared their results with five open-source projects. They concluded that the techniques tend to have an excellent performance when applied to small-scale projects. They identify some industrial issues that lead to performance degradation compared with open-source projects. They also surveyed 25 Huawei practitioners about their willingness to adopt the IRBL approaches analyzed. Most are willing to adopt IRBL tools in bug localization activities, but existing techniques have not met their desired extent. 

In our previous work \cite{medeiros2020}, we applied techniques for grouping crash reports and finding buggy code in three large-scale web-based systems. We found results consistent with those reported by other researchers. For instance, we successfully identified buggy classes with recall varying from 61.4\% to 77.3\%, considering the top 1,  3, 5 and 10 suspicious buggy files identified and ranked by our approach. We also found that 80\% of the changed methods from closed bug fix issues appeared in related stack traces of crash report groups. However, we analyzed crash reports and bugs that had already been fixed previously --- a retrospective study. 

In this paper, we apply our approach to existing crash reports to identify if they can help the development team in bug-fixing tasks on the same set of projects. Furthermore, we analyze the results from the developers’ point of view, which, as far as we know, has not been reported in the literature. We identified that 96\% of the files changed to fix the bug was one of the suspicious files suggested by the approach, which is better than the values obtained in our previous study\cite{medeiros2020}, which varied between 61.4\% and 77.3\%. The Mean Average Precision of 86\% was better than our first study, which ranged between 35.7\% and 45.9\%. According to the developers’ answers, the suspicious files ranking helped to fix the bug for 78.1\% of the tasks. Similarly, the list of suspicious methods helped by 65.9\%. During the weekly meetings, we also got reports that file ranks and suspicious methods help even when the bug is not present in them. In addition, other differences from this work are the list of criteria used by the teams to prioritize the log groups to be analyzed and what extra information provided by the approach helped fix the bugs.

\section{Conclusion} \label{sec:conclusion}

In this paper, we reported our experience over 18 months of using an approach to group crash reports by stack trace and indicate buggy components to reduce the cognitive effort to identify the root causes of a system crash. We applied it to open bug-fix tasks for development teams responsible for three non-trivial Java web systems. Our results bring new evidence that the approach is effective in locating the origin of defects. 

We observed that developers' main criteria for prioritizing a group were critical requirements, log group size, and difficulty reproducing the error. Our approach obtained an average recall greater than 90\% and a mean average precision above 82\%. Additionally, developers reported that lists of suspicious files and methods help even when the bug is not present in any code in the ranking. We also verified that stack trace and URIs were more helpful information during bug-fixing.

The development teams continue using our approach to correct bugs related to the stack trace groupings. Our subsequent work will investigate how to provide information that facilitates choosing error groups that should be prioritized and how we can improve the reproducibility of the bugs using information from crash reports.

\begin{acks}
We thank all the developers, team leaders, and managers from STI/UFRN  who contributed to our study. This work is partially supported by INES (www.ines.org.br), CNPq grant 465614/2014-0, CAPES grant 88887.136410/2017-00, FACEPE grants APQ-0399-1.03/17, PRONEX APQ/ 0388-1.03/14, and CNPq grant 425211/2018-5.
\end{acks}

\bibliographystyle{ACM-Reference-Format}
\bibliography{sample-base}

\end{document}